\documentclass[twocolumn,superscriptaddress,showpacs,prb]{revtex4}
\usepackage{graphicx}
\usepackage[usenames]{color}
\usepackage{amsmath}

\newcommand{\Ang}{{\rm \AA}}

\newcommand{\be}{\begin{equation}}
\newcommand{\ee}{\end{equation}}
\newcommand{\ben}{\begin{eqnarray}}
\newcommand{\een}{\end{eqnarray}}
\newcommand{\beq}{\begin{equation}}
\newcommand{\eeq}{\end{equation}}

\newcommand{\ds}{\displaystyle}

\begin{document}


\title{Numerical studies of confined states in rotated bilayers of graphene}

\author{G. Trambly de Laissardi\`ere}
\email{guy.trambly@u-cergy.fr}
\affiliation{Laboratoire de Physique Th\'eorique et Mod\'elisation, Universit\'e de Cergy-Pontoise - CNRS,
F-95302 Cergy-Pontoise Cedex, France.}
\author{D. Mayou}
\email{didier.mayou@grenoble.cnrs.fr}
\author{L. Magaud}
\email{laurence.magaud@grenoble.cnrs.fr}
\affiliation{Institut N\'eel, CNRS - Universit\'e Joseph Fourier, F-38042 Grenoble, France}

\date{\today}


\begin{abstract}
Rotated graphene multilayers form a new class of graphene related systems with electronic 
properties that drastically depend on the rotation angles. It has been shown that bilayers  
behave like two isolated graphene planes for large rotation angles.  
For smaller angles, states in the Dirac cones belonging to the two layers interact 
resulting in the appearance of two van Hove singularities. States become localised 
as the rotation angle decreases and the two van Hove singularities merge into one peak 
at the Dirac energy. Here we go further and consider bilayers with very small rotation angles.  
In this case, well defined regions of AA stacking exist in the bilayer supercell and we show 
that states are confined in these regions for energies in the [$-\gamma_t$, $+\gamma_t$] 
range with $\gamma_t$ the interplane  mean interaction.  As a consequence, the local densities 
of states show discrete peaks for energies different from the Dirac energy. 
\end{abstract}



\maketitle

\section{Introduction}

Graphene exceptional electronic properties are predicted for an isolated layer of C atoms 
arranged on a pristine honeycomb lattice.\cite{Wallace47,Castro09_RevModPhys}  
Stacking several layers on top of each other, may affect the original electronic structure. 
Indeed AB or Bernal stacking (such as what is found in graphite)  destroys both the linear 
dispersion and the chirality properties even in a 
bilayer.\cite{Latil06, MacdoABC, Varchon_prb08, Ohta06, Coletti10, Brihuega08}  
Many experimental realisations of graphene lead to the formation of multilayers 
(on SiC \cite{Ohta06, Coletti10, Brihuega08, Hass06, Hass08, Emtsev08, Hass_prl08, Sprinkle09, Hicks11}  
but also on metal surfaces such as Ni\cite{Andrei_Ni11}  
and in exfoliated flakes\cite{ Andrei_graphite10}) 
where interactions between successive layers play a crucial role. 
While on the Si face of SiC, multilayers have an AB stacking and do not show graphene properties,\cite{Varchon_prb08,Ohta06,Coletti10,Brihuega08} 
on the C-face, multilayers have been shown  to present graphene like properties even when they 
involve a large number of C planes. ARPES,\cite{Emtsev08,Sprinkle09,Hicks11} STM,\cite{Miller09} 
transport,\cite{Berger06} and optical transitions\cite{Sadowski06} indeed 
show properties characteristic of a linear graphene like dispersion. 
These multilayers are rotated with respect to each other and the  
rotations show up as moir\'e  patterns on STM images.\cite{Hass_prl08, Varchon_stm08, Rong93}
This apparent controversy -thick multilayers exhibiting graphene properties- was partially 
solved recently when different theoretical 
approaches\cite{Dossantos07,CastroNeto_condmat, Latil07, Hass_prl08, Shallcross08, Mele, Mele2, Suarez, Bistritzer_PNAS, Bistritzer2, Trambly10}  
showed that rotated multilayers are decoupled, at least for large rotation angles. 
Going further, theory predicts the existence of three domains that in fact correspond to two regimes: 
for large rotation angles, $\theta >15^{\mathrm o}$, 
the layers are decoupled and behave as a collection of isolated graphene layers. For intermediate angles,
$2^\mathrm{o}<\theta<15^\mathrm{o}$, the dispersion remains linear but  the velocity is renormalised. 
For these large and intermediate angles, Dirac cones persist and the interaction between rotated layers 
is a perturbation. What happens at the smallest values of $\theta$ is more puzzling. 
As already shown by different theoretical groups, for the lowest $\theta$, 
flat bands appear and result in electronic localization. 

From the experimental point of view, while some experiments with graphene on SiC show none of 
the effects predicted by theory (no renormalisation nor van Hove singularities),\cite{Hicks11} 
others\cite{Veuillen} do find them. 
Furthermore, Landau Level (LL) Scanning Tunneling Spectroscopy (STS) 
gave results in close agreement with theory and the three domains were observed for CVD graphene grown on 
Ni.\cite{Andrei_Ni11}

Here we go further and show that in the very small angle regime, 
confinement occurs and states become confined in the AA (all atoms 
on top of each other) region. Furthermore, confinement effects lead to the formation 
of sharp peaks in the local density of states in an energy window that is fixed by the 
interplane mean interaction. 
Thus, this apparently simple system -a graphene bilayer-  presents a complex behavior ranging 
from ballistic to confined electrons when the rotation angle is changed. 

Tackling very small rotation angles - smaller than $1^\mathrm{o}$ - 
means handling very large cells that involve a huge number of atoms - more than 60\,000 -. 
This cannot be done from ab initio calculations and we developped a tight binding scheme 
\cite{Trambly10} using only $p_z$ orbitals to do so. Rotated multilayers are described in part II, 
the method used is outlined in the third part and described in more details in the Appendix A. 
Part IV gives the variation of the van Hove singularity position with the rotation angle 
and parts V and VI describes the electronic structure of graphene rotated bilayers 
in the case of very small rotations angles.

\section{Rotationaly stacked commensurate bilayers}

In the following we consider two graphene layers rotated in the plane 
by an angle $\theta$. We start from an AA bilayer and choose the rotation origin O  at an atomic site. 
Since we want to perform ab initio and Tight Binding calculations, we need a periodic system.
A commensurate structure can be defined if the rotation changes a lattice vector 
${\overrightarrow{OB}} ~(m,n)$ to
${\overrightarrow{OB'}}~(n,m)$ with $n$, $m$ the coordinates with respect to the basis vectors 
$\vec a_1$ $(\sqrt{3}/2,-1/2)$  
and $\vec a_2$ $(\sqrt{3}/2,1/2)$. 
The rotation angle is then defined as follows:
\begin{equation}
\cos\theta = \frac{n^2+4nm+m^2}{2(n^2+nm+m^2)}
\label{eq:Comm3}
\end{equation}

and the commensurate cell vectors correspond to :
\begin{equation}
\vec t = {\overrightarrow{OB'}} = n \vec a_1 + m \vec a_2\,,~ \\
\vec{t'} = -m \vec a_1 + (n+m) \vec a_2.
\label{eq:Comm4}
\end{equation}

The commensurate unit cell contains $N = 4(n^2+nm+m^2)$ atoms. As we have already shown,\cite{Trambly10} 
the rotation angle is a good parameter to describe the system but the number of atoms is not since cells 
of equivalent size can be found for different angles. Indeed, large cells can be obtained for $\theta \sim 0$  
- large $n$ and $m$ and small $|m-n|$ - but  also  for large angles $\theta \sim 30^\mathrm{o}$ 
- then $|m-n|$ is large\,-.\cite{Campanera}  A good way to name the rotated layers is to give the 
($n,m$) couple defined above, which is what we use in the present paper.
$\theta = 60^\mathrm{o}$ is the perfect AB stacking. $\theta$ close to $60^\mathrm{o}$ is 
obtained for $n=1$ ($m=1$) and large $m$ ($n$), it is equivalent in our calculations 
to  $\theta \sim 0^\mathrm{o}$.

\section{Methods}

Our approach combines {\it ab initio} and tight binding calculations. First principle approach can tackle cells 
with up to $\sim 1000$ atoms. The tight-binding (TB) scheme allows electronic structure calculations by 
diagonalisation in reciprocal space for crystal structures 
containing up to 59\,644 atoms in a unit cell 
- corresponding to $\theta=0.469 ^\mathrm{o} $ and the ($70,71$) bilayer -. 
By the using recursion technique in real space, very large structures up to several millions 
- 3\,006\,004 for the (500,501) with $\theta = 0.067^\mathrm{o}$ - of atoms are studied. 
The tight binding scheme is described in Ref. \onlinecite{Trambly10} and in the Appendix; 
it is restricted to $p_z$ orbitals.  
It was developped from ab initio calculations and it gives a good description of the rotated bilayers, 
for small angles as well as large ones but also for trilayers, graphene and graphite.  
The TB parameters were fitted to reproduce the velocity of monolayer graphene and the interlayer mean 
interaction $\gamma_t= 0.34$\,eV (see Appendix). Flat layers were used. We performed an ab initio calculation 
for the (6,7) bilayer allowing all the atoms to relax and found only minor changes: 
the layer corrugation is small and the effect on the band structure is very small.\cite{LDA}
Velocities are calculated from the slope of the dispersions at the Dirac point along $\Gamma$K.

\section{Van Hove singularities}

The rotated bilayer has already been the subject of many theoretical developments either analytical or 
numerical.\cite{Dossantos07, CastroNeto_condmat, Shallcross08, Mele, Mele2, Latil07, Hass_prl08, Suarez, Bistritzer_PNAS, Bistritzer2, Trambly10}
The main results are only briefly recalled in the following since we want to focuss on the very 
small rotation case and the confinement effect that results. These theoretical predictions are in 
very good agreement with Scanning Tunneling Spectroscopy 
experiments on CVD graphene grown on nickel.\cite{Andrei_Ni11} 
Properties are symmetric with respect to $\theta=30^\mathrm{o}$: 
In our calculations, the behavior is the same for angles decreasing to $0^\mathrm{o}$  or increasing 
to $60^\mathrm{o}$.

\begin{figure} 
~~~~~~~\includegraphics[width=5cm]{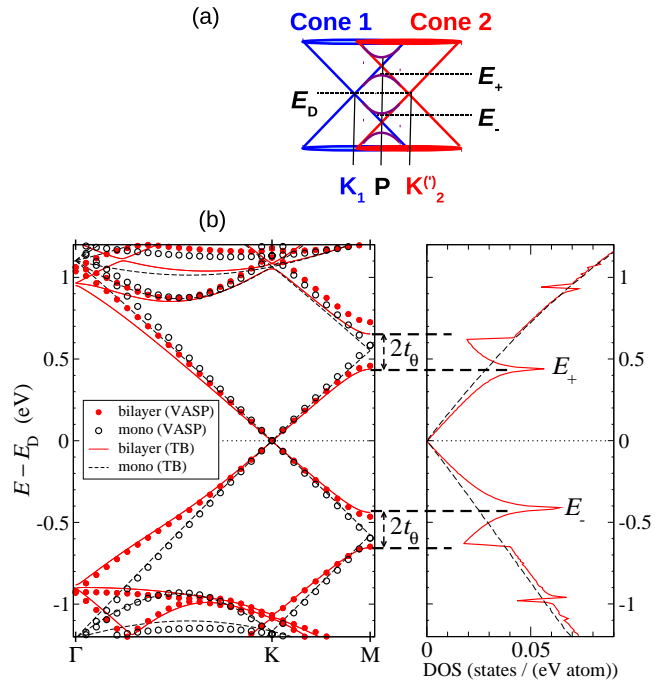}

\includegraphics[width=8.5cm]{Fig_n4m5_bnds_DOS_VASP_TB_v32.eps}
\caption{(color on line)
Van Hove singularities of a (4,5) bilayer. (a) schematic drawing, (b) (4,5) band structure and total DOS.
}
\label{Fig1} 
\end{figure} 

For large angles, $\theta> 15^\mathrm{o}$,   
the rotated bilayer behaves like two isolated graphene layers. 
For intermediate angles, $2^\mathrm{o}<\theta<15^\mathrm{o}$,  
the dispersion remains linear but the velocity is strongly renormalised. 
In this regime, one important feature is the appearance of van Hove singularities on both sides of the Dirac energy.
In the following, we consider that layer 2 is rotated by an angle $\theta$ with respect to layer 1. 
Layer 2 Brillouin zone is then rotated by the same $\theta$ with respect to layer 1 Brillouin zone. 
The closest Dirac cones belonging to layer 1 and 2 intersect at a point P which is at mid-distance between 
the $K_1$ and $K_2$  points of the 1x1 Brillouin zones (see Fig.\,\ref{Fig1}a). 
P is backfolded to point M of 
the supercell brillouin zone. Interaction of the states with similar energies opens two gaps at the saddle point 
M (Fig.\,\ref{Fig1}b). Flat bands have a drastic effect on the electronic structure of a 
two-dimensionnal systems and appear as van Hove singularities on each side of the Dirac energy 
 (Fig.\,\ref{Fig1}b). In fact, each gap creates two van Hove singularities: 
Singularities closest to the Dirac point appear as peaks $E_+$ and $E_-$ in Fig.\,\ref{Fig1}b, 
they are followed by a dip and a sharp increase of the DOS that corresponds to the flat bands 
on the other side of the gaps. 
The gap width and then the energy difference between the steep increase in the DOS and the peak 
are equal to $2t_\theta$ with $t_\theta$ the mean interaction between states related to the two layers 
at the intersection between the Dirac cones, as defined by Ref.\,\onlinecite{CastroNeto_condmat}.  

\begin{figure}
\includegraphics[width=8cm]{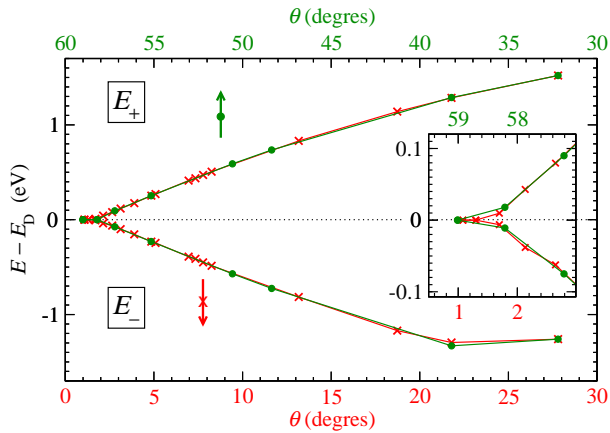}
\caption{(color on line) 
Energies of the van Hove singularities 
(the two peaks closest to $E_{D}$) in ($n,m$)  
bilayers versus $\theta$).
Lines are guide for the eyes. 
} 
\label{Fig_E_VanHove}
\end{figure}

\begin{figure}
\includegraphics[width=6cm]{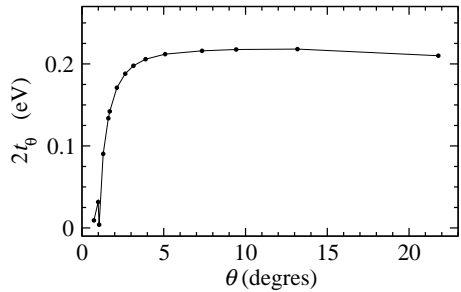}
\caption{
Gap at P (Point M of the supercell) as a function of $\theta$. Line is a guide for the eyes.
}
\label{Fig_gap}
\end{figure}

In Fig.\,\ref{Fig_E_VanHove},  the energies, $E_+$ and $E_-$ of the closest peaks to the Dirac energy are 
plotted versus $\theta$. 
For intermediated angle values, 
$2 <  \theta  < 15^o$, $E_+$ and $E_-$ are linear with respect to $\theta$.

As the rotation angle decreases, the intersections of the two Dirac cones belonging to the two graphene layers 
come closer to the Dirac 
energy (Fig.\,\ref{Fig1}).  
The intersection point P is at mid-distance between $K_1$ and $K_2$. 
$K_1K_2$ distance decreases linearly with $\theta$. Since in this regime the dispersion is linear and the gap at P is 
roughly constant (Fig.\,\ref{Fig_gap}),  the position of the van Hove singularities then varies linearly with $\theta$.

For small angle values, $\theta \sim 2^\mathrm{o}$, $E_+$ and $E_-$ merge and form a sharp peak 
at the Dirac energy $E_\mathrm{D}$.

As shown in our previous article,\cite{Trambly10} 
this sharp peak corresponds to states strongly localised in AA stacking regions of the moir\'e. 
Recently Lopes Dos Santos et al.\cite{CastroNeto_condmat}
extended their treatment based on the continuum model to smaller angles  and confirmed the localisation 
of the states in the AA regions of the moir\'e pattern.
 
\section{Confinement in AA regions}

\begin{figure}[ht]
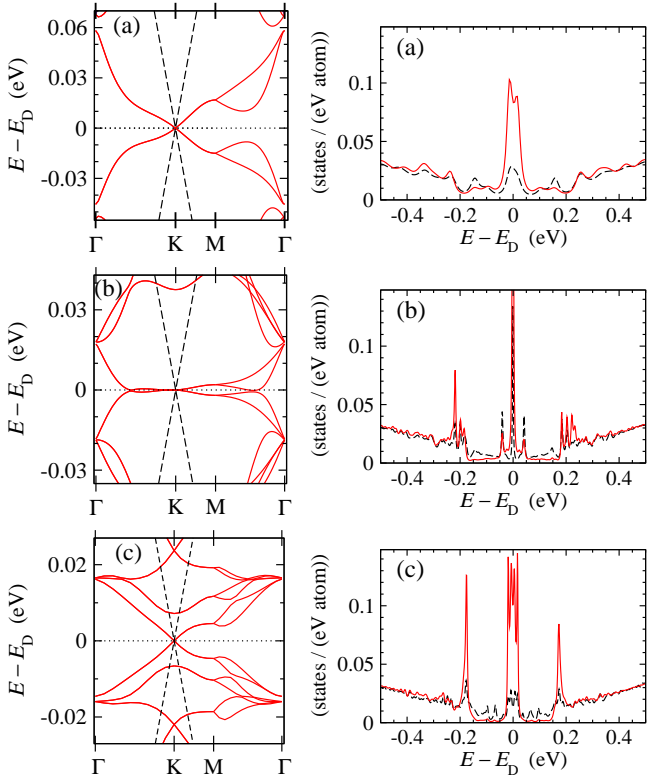

\begin{center}

\includegraphics[width=3.8cm]{Fig_n19m20_bndsGKMG_ED0.eps}
~\includegraphics[width=4.5cm]{Fig_n19m20_DOS_t_AA_ED0.eps}

\vskip .1cm
\includegraphics[width=3.8cm]{Fig_n31m32_bndsGKMG_ED0.eps}
~\includegraphics[width=4.5cm]{Fig_n31m32_DOS_t_AA_ED0.eps}

\vskip .1cm
\includegraphics[width=3.8cm]{Fig_n45m46_bndsGKMG_ED0.eps}
~\includegraphics[width=4.5cm]{Fig_n45m46_DOS_t_AA_ED0.eps}

\end{center}
\caption{(color on line) 
Band structures and DOS for bilayers with rotation angles that lead to van Hove singularities 
close to the limit of the linear variation. 
(a) (19,20), $\theta = 1.67^\mathrm{o}$, 
(b) (31,32), $\theta = 1.05^\mathrm{o}$, 
(c) (45,46), $\theta = 0.73^\mathrm{o}$. 
Left: Dispersion of monolayer graphene is shown in black dashed line for comparison.
Right: (red line) local DOS at the center of AA zone, (black dashed line) total DOS.
}
\label{Fig_limvH}
\end{figure}

It is interesting to look at the band structures of bilayers with small angles -small enough to have 
van Hove singularities at the limit of the 
linear variation (Fig.\,\ref{Fig_limvH})-. 
(19,20) bilayer is at the edge of the linear regime: the band structure still shows a linear dispersion and 
van Hove singularities. 
For the (31,32) bilayer, bands are flat and velocity is equal to zero\cite{Trambly10} 
but in the case of the (45,46) bilayer, 
the peak at $E_\mathrm{D}$  is split in two and the velocity at the Dirac point increases again 
and is equal to $\sim 20$\% of the graphene velocity. 
This is coherent with the increase of velocity between magic angles found by Bistritzer 
and MacDonald \cite{Bistritzer_PNAS} (section \ref{SecMagicAngles}).

It is important to notice that near the Dirac point one observes the existence of two bands that  
disperse linearly. These bands have a twofold degeneracy (Dirac cones from the two layers).  
The 4 bands contributes to the central peak of the total density of states. 
The weight of the eigenstates close to the Dirac energy is strong in the AA zone as shown by 
diagonalization.\cite{Trambly10,note}

\begin{figure}
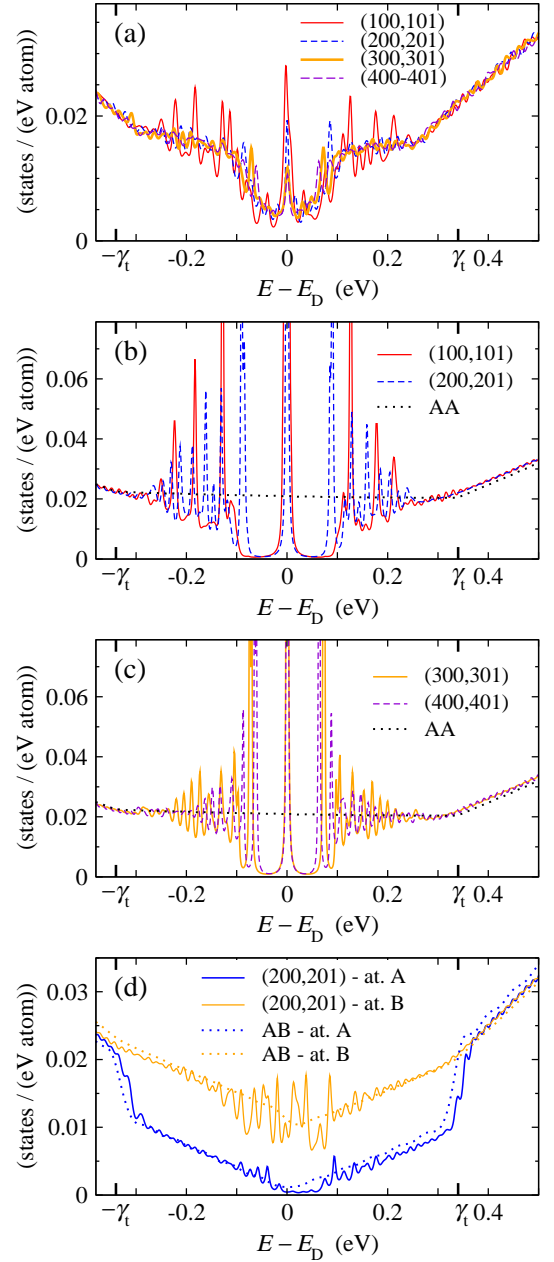


\includegraphics[width=7cm]{FigDOSt_100_200_300_400_ED0.eps}

\vskip .1cm
\includegraphics[width=7cm]{Fig_LDOSAA_100_200_ED0.eps}

\vskip .1cm
\includegraphics[width=7cm]{Fig_LDOSAA_300_400_ED0.eps}

\vskip .1cm
\includegraphics[width=7cm]{Fig_LDOSAB_AB_200-1_ED0.eps}

\caption{(color on line) 
TB DOS in bilayers: 
(a) Total DOS.
Local DOS: (b) (c) on atom at the center of an AA zone,
(d) on atom at the center of an AB zone 
(At. A: with an atom directly on top or below and 
at. B: without an atom on top or below).
($n$,$m$) bilayers: values
(100,101) $\theta = 0.33^\mathrm{o}$,
(200,201) $\theta = 0.16^\mathrm{o}$
and
(300,301) $\theta = 0.11^\mathrm{o}$.
(400,401) $\theta = 0.08^\mathrm{o}$.
}
\label{Fig_DOS_PetitsAngles}
\end{figure}

\begin{figure}
\includegraphics[width=5cm]{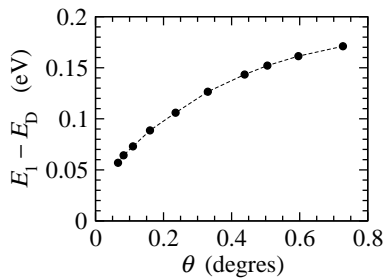}
\caption{
Variation of the energy of the first peak ($E_1$) away from the Dirac point of the local DOS 
at the center of AA region as a function of $\theta$.
}
\label{Fig_DE_AA}
\end{figure}

\begin{figure}
\includegraphics[width=6cm, angle=-90]{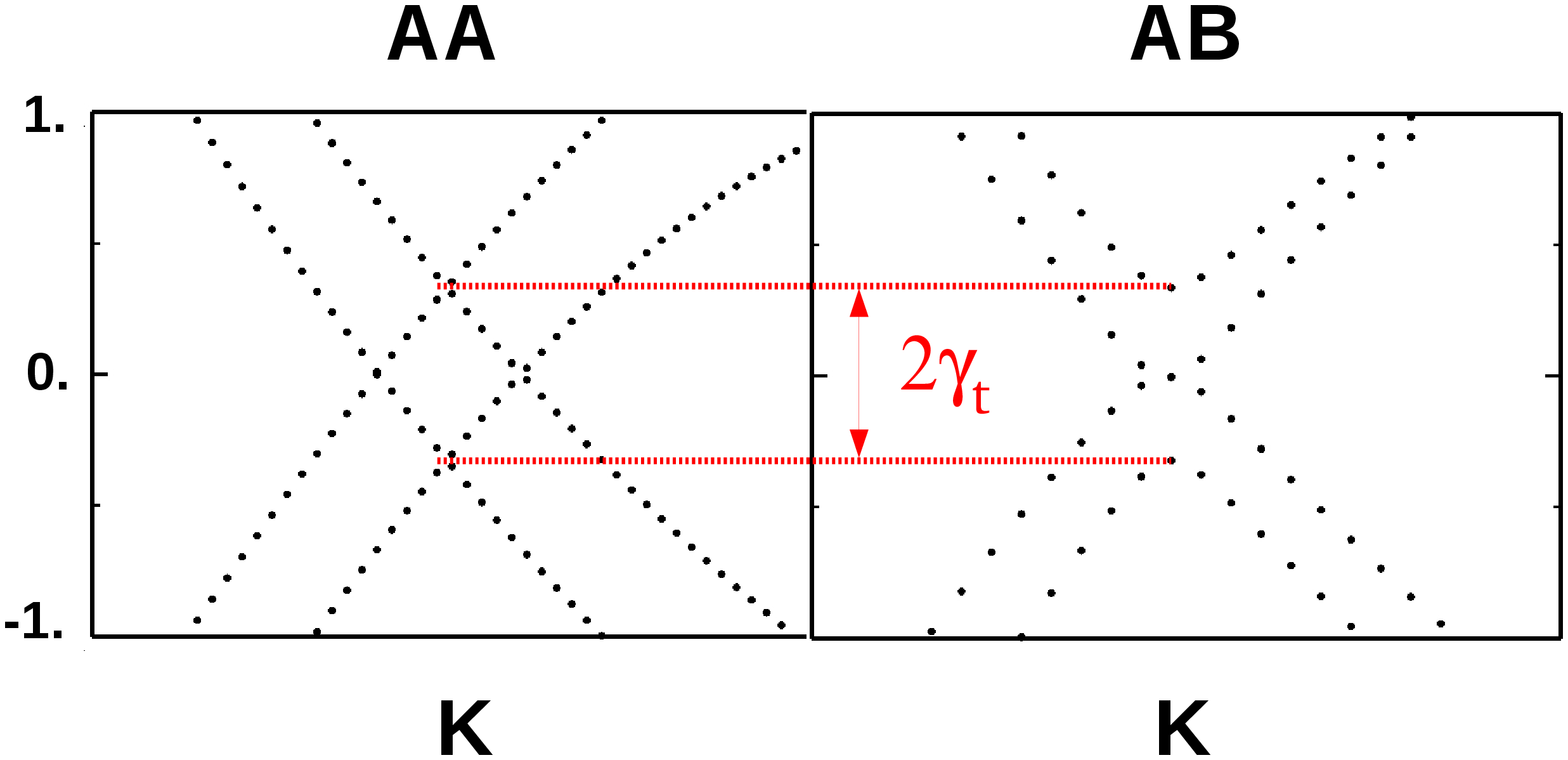}
\vskip -.5cm
\caption{(color on line)
Band structure of an AA and an AB stacked bilayer in the vicinity of the Dirac Energy.} 
\label{Fig_AA_AB}
\end{figure}

The analysis of the density of states, shows that a specific regime occurs 
for very small angles, typically $\theta \le 1^\mathrm{o}$. 
The total DOS (Fig.\,\ref{Fig_DOS_PetitsAngles}a) 
contains sharp peaks at energies different from the Dirac energy.
Figs.\,\ref{Fig_DOS_PetitsAngles}b and \ref{Fig_DOS_PetitsAngles}c 
show the local DOS at the center of an AA stacked zone for decreasing $\theta$, 
and Fig.\,\ref{Fig_DOS_PetitsAngles}d the local DOS at the center of an AB stacked zone.

For so small rotation angles, the moir\'e pattern presents well defined regions of AA and AB stacking. 
In the center of these regions, the local DOS follows the AA or AB DOS (dotted lines in 
Figs.\,\ref{Fig_DOS_PetitsAngles}b, c and d), except for energies close to the Dirac energy -. 
Indeed according to Fig.\,\ref{Fig_DOS_PetitsAngles}  peaks appear in an energy window defined by the 
mean interplane interaction  $\gamma_t$.
DOS in intermediate regions (not shown) 
are very similar to DOS in AB regions and do not show large peak.

These results are consistent with confinement effect that can explain the discrete character of 
the spectrum in the $[-\gamma_t,+\gamma_t]$ range. Indeed, in large moir\'e patterns, AA regions seem isolated 
by AB and intermediate regions. From Fig.\,\ref{Fig_DOS_PetitsAngles}d, we can say that confinement 
in AB regions is much less efficient than in AA regions. In a nanostructure, confinement results in peak 
in the DOS whose position depends on the size of the nanostructure. 
Here the size of the AA region is inversely proportional  
to the rotation angle $\theta$. 
In Fig.\,\ref{Fig_DE_AA} we plot the energy of the first peak away from the Dirac energy  
as a function of $\theta$. As expected this energy decreases when the size of the AA region increases, 
which is when the angle  $\theta$ decreases.

In the energy range  $[-\gamma_t,+\gamma_t]$, the symmetry of the wavefunction
is different in an AA or AB bilayer 
(Fig.\,\ref{Fig_AA_AB}, see also the DOS in the appendix Fig.\,\ref{Fig_DOS_graphene_LMTO_TB}b). 
Because of the different symmetry of the wavefunctions, matching is difficult which could explain that  
the AA stacked region is isolated. We note also  that in this energy range the coupling between electrons 
and holes which have opposite velocities could favor localization. 
On the contrary out of the energy range  $[-\gamma_t,+\gamma_t]$, 
the mixing between the hole states of one plane and the electron states 
of the other plane is weak. Indeed the coupling occurs between electron states  
of one plane and hole states of the other plane that have the same wave vector 
but have a difference in energy that is much larger than $\gamma_t$. 
Therefore outside of the energy range  $[-\gamma_t,+\gamma_t]$ the mixing 
of states of the two planes occurs separately between the electron states of the two planes and between 
the hole states of the two planes. We suggest that this absence of mixing between electrons and holes 
which have opposite velocities suppresses the localization.

\begin{figure} 
\includegraphics[width=8.5cm]{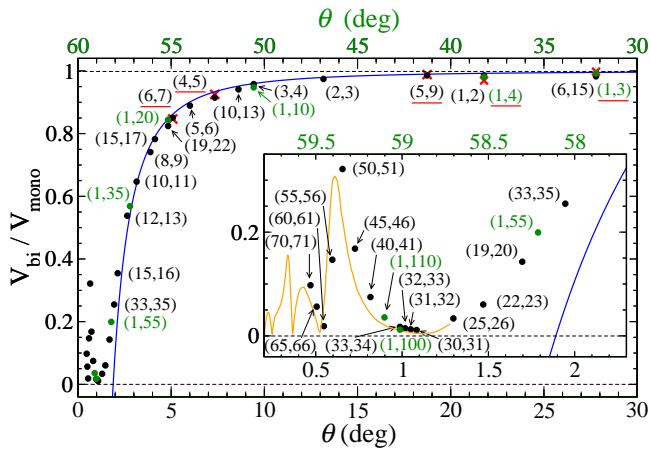}
\caption{(color on line)
Velocity of a bilayer divided by the velocity of a monolayer, at the Dirac point, as a function of $\theta$. 
The insert shows a zoom on very small angles and on angles close to $60^o$. 
Black/green point: TB calculation,  
red cross: ab initio calculation, blue line: law proposed by Lopes dos Santos et al.,\cite{Dossantos07} 
orange line: model proposed by Bistritzer and MacDonald\cite{Bistritzer_PNAS} 
for very small angles. 
The latter line is rescaled to obtain a first minimum of the velocity at $1.13^\mathrm{o}$ 
as in TB calculations (see text).
Ab initio calculations and most of TB calculations for $\theta > 1.05^o$ 
are from Ref. \onlinecite{Trambly10}.
} 
\label{Fig_vitesse} 
\end{figure}

\section{Magic angles}
\label{SecMagicAngles}

As discussed previously, at very small angles, the velocity of rotated bilayers 
is no longer monotonic but shows minima and maxima (Fig.\,\ref{Fig_vitesse}). 
Minima correspond to a flat band  with zero velocity at the Dirac energy. Bistritzer 
and MacDonald in 
Ref. \onlinecite{Bistritzer_PNAS} found that the velocity is equal to zero for special magic 
angles  $\theta_n$. The values in degrees are  
$\theta_1 =1.05$, $\theta_2=0.5$, $\theta_3=0.35$, $\theta_4=0.24$, $\theta_5=0.2$. 
We note that this series is simply given by $\theta_n=1.05/n$. 
In the present calculations we recover by a full tight-binding calculation 
the first two angles $n=1$ and $n=2$ (see Fig.\,\ref{Fig_vitesse}) which 
are given by $\theta_n=1.13/n$ in good accordance with the more simplified model \cite{angles}
of Ref.\,\onlinecite{Bistritzer_PNAS}.

Below we propose a simple heuristic argument which relates these magic angles 
to quantization conditions. At these magic angles the bands become flat  which 
means that the states of the different  AA zones are decoupled and thus confined in the AA zones. 
In the following, we use a hard wall model to describe the confinement effect. 
The confinement condition is typically:
\begin{equation}
kd +2\Phi =2n \pi,
\label{eq:Comm4}
\end{equation}
where $k$ is the wavevector counted from the K point of the Dirac wave in the AA zone, 
$\frac{d}{2}$ is the typical size of the AA zone and $\Phi$ is the phase accumulated 
at the reflexion at the boundary of the AA zone. We assume that this phase can be integrated in an efficient 
size of the AA zone so that the quantization condition finally gives:
\begin{equation}
xkD =2n \pi,
\label{eq:Comm5}
\end{equation}
where $D$ is the period of the moir\'e and $x$ is a numerical coefficient of the order of 
1  such that $xD/2$ is the effective size of the AA zone. 
This expression can be cast in terms of the parameters of the present study. Indeed  
the period $D$ of the moir\'e expressed in nanometers and the angle $\theta$ of the rotation 
expressed in degrees are related by  $D(\mathrm{nm} )=\frac{14}{\theta(\mathrm{deg}) }$ for small angles. 
In addition, we have the relation $\hbar v_F k=\gamma_t$ between the parameter $\gamma_t$ and the wavevector 
$k$ at the Dirac energy in the AA zone. Then the quantization relation leads to:
\begin{equation}
\frac{3.4 x\gamma_t (\mathrm{eV})}{n} \sim \theta (\mathrm{deg}).
\label{eq:Comm6}
\end{equation}
For the present value of the 
parameter $\gamma_t =0.34$\,eV  and with a reasonable value of the parameter $x \sim 0.95$  
this relation leads exactly to the magic angles 
$\theta (deg)=1.13/n$ found here.  This value is close to the series  $1.05/n$ found in 
Ref. \onlinecite{Bistritzer_PNAS}. 

Why the simple hardwall argument discussed here can describe the magic angles series is not obvious.  
Indeed Bistritzer and MacDonald\cite{Bistritzer_PNAS} 
used an interplane interaction hamiltonian with a sinusoidal variation in space. 
For free electrons a sinusoidal interaction potential would probably not lead to  
the localization with hard wall conditions.  Yet as discussed above, 
the localization in the energy range $[-\gamma_t,+\gamma_t]$ that is found in this 
work suggest that the spinor character of the electronic wavefunction in  graphene  
leads to  more complex localization phenomena.

\section{Conclusion}

Thanks to the tight binding scheme that we developped, we studied the electronic structure of 
rotated bilayers in the case of very small rotation angles, which is beyond what was recently proposed. 
We showed that states are confined in regions of well defined AA stacking and as a consequence that 
the LDOS shows discrete peaks not only at the Dirac energy but in the $[-\gamma_t,+\gamma_t]$ range 
with $\gamma_t$  
the mean interlayer hopping integral. Very large moir\'e patterns 
(corresponding to angles close to $1^{\mathrm{o}}$) 
have already been observed experimentally. This system could then offer the possibility to study the physics of 
a regular array of weakly interacting dots.
To conclude we note that the rotated bilayers of graphene present analogies with quasicrystals 
and approximant crystals that have also large unit cells and tend to confine electrons 
(Ref. \onlinecite{Berger93} and references within).
In the context of quasicrystals this localization by large unit cells leads 
to unique transport properties.\cite{Triozon02,Trambly06} 
This  suggests that rotated graphene bilayers 
with large unit cells could have original transport properties.

\section*{Acknowledgments}

The authors wish to thank P. Mallet and J.-Y. Veuillen for fruitful discussions. 
The computations have been performed at the
Service Informatique Recherche (SIR),
Universit\'e de Cergy-Pontoise and on the CIMENT/PHYNUM project calculation center.
We thank Y. Costes and David Domergue, Universit\'e de Cergy-Pontoise, 
and Dan Calugaru, Institut N\'eel,
for computing assistance. We acknowledge financial support from RTRA (DISPOGRAPH project) 
and the french national agency (ANR P3N/NANOSIM-GRAPHENE project).

\appendix
\section{Tight Binding scheme for bilayers}

\begin{figure}
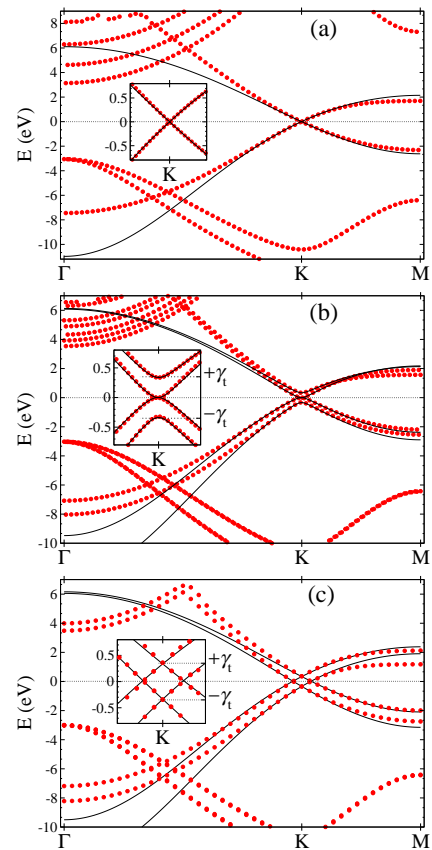

\includegraphics[width=5.5cm]{Fig_graphene_VASP_TB_v2.eps}

\vskip .1cm
\includegraphics[width=5.5cm]{Fig_AB_VASP_TB_v2.eps}

\vskip .1cm
\includegraphics[width=5.5cm]{Fig_AA_abinitio_TB.eps}

\caption{(color on line) Band structure $E(\vec k)$ calculated from {\it ab initio} VASP and 
TB:
(a) graphene, (b) AB bilayer, (c) AA bilayer. $\gamma_t=0.34\,eV$. }
\label{Fig_graph_mono_AA_AB}
\end{figure}

\begin{figure}
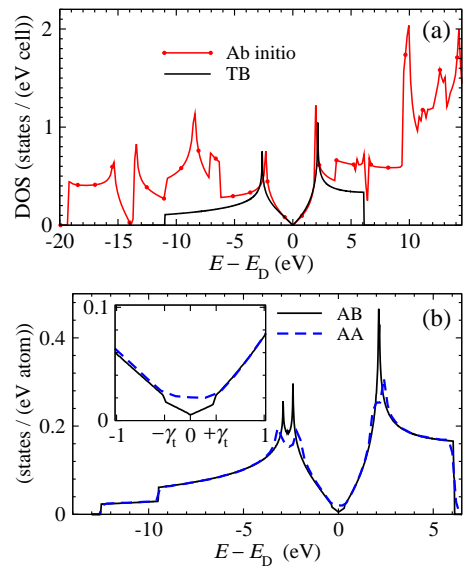

\includegraphics[width=6cm]{Fig_DOSgraphene_TB_LMTO.eps}

\vskip .1cm
\includegraphics[width=6cm]{Fig_DOS_AA_AB.eps}

\caption{(color on line) 
Density of states: 
(a) ab initio (Linear Muffin Tin Orbitals \cite{LMTO}  and TB total DOS in graphene, 
(b) TB total DOS in AB (Bernal) and AA bilayers.
[Insert: total DOS arround the $E_{\rm D}=0$. $\gamma_t=0.34\,eV$]
\label{Fig_DOS_graphene_LMTO_TB}
}
\end{figure}

\begin{figure}
\includegraphics[width=6.5cm]{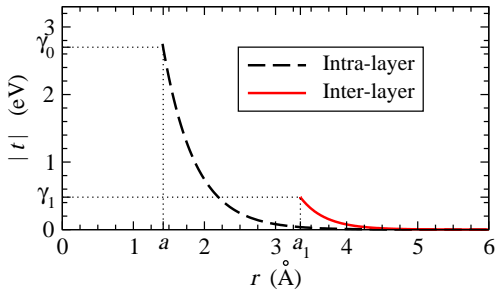}
\caption{Coupling term $t$ between two $p_z$ orbitals versus the distance $r$: 
(full line) in two layers (inter-layer coupling); (dashed line) in the same layer (intra-layer coupling). 
Here, the smooth cut-off function $F_c$ is not taken into account ({\it ie} 
large  $r_c$ value in (\ref{Eq_cutoffFunction})).}
\label{Fig_couplageIntraInter}
\end{figure}

In this appendix we present in details the 
tight binding (TB) scheme that we have used in our previous work\cite{Trambly10} and in the 
present results. 
It reproduces
the {\it ab initio} calculations 
of the electronic states for energies within $\pm 2$\,eV of $E_{\rm D}$ whatever the rotation 
angle is (Fig. \ref{Fig_graph_mono_AA_AB}, \ref{Fig_DOS_graphene_LMTO_TB} and 
Ref. \onlinecite{Trambly10}).
Here $E_{\rm D}$ is set to zero.

Only $p_z$ orbitals are taken into account since we are interested in what happens at the Fermi level. 
Since the planes are rotated, neighbours are not on top of each other 
(as it is the case in the Bernal AB stacking). Interlayer interactions are then not restricted to 
$pp\sigma$ terms but some  $pp\pi$ terms have also to be introduced. 
The Hamiltonian has the form :
\ben
H &=& \sum_{i} \epsilon_i \, |i\rangle \langle i|  
~+~  \sum_{<i,j>} t_{ij} \, |i\rangle \langle j| ,
\label{Eq_hamilt}
\een
where  $|i\rangle$ is the $p_z$ orbital located at $\vec r_i$, and $\langle i,j\rangle$ is 
the sum on index $i$ and $j$ with $i\ne j$.
The coupling matrix element, $t_{ij}$, 
between two $p_z$ orbitals located at $\vec r_i$ and $\vec r_j$ 
is,\cite{SlaterKoster54}
\ben
t_{ij} 
~=~   n^2 V_{pp\sigma}(r_{ij}) ~+~ (1 - n^2) V_{pp\pi}(r_{ij}),
\een
where $n$ is the direction cosine of 
$\vec r_{ij} = \vec r_j - \vec r_i$ along $(Oz)$ axis 
and $r_{ij}$ the distance $r_{ij}$ between the orbitals,
\ben
n ~=~ \frac{z_{ij}}{r_{ij}} ~~~{\rm and}~~r_{ij} = ||\vec r_{ij}|| .
\een
$z_{ij}$ is the coordinate of $\vec r_{ij}$ along $(Oz)$. It is either equal to zero or 
to a constant because the two graphene layers have been kept flat in our model.\cite{relax}
We use the same following dependance on distance of the Slater and Koster parameters:
\begin{eqnarray}
V_{pp\pi}(r_{ij})    &=&-\gamma_0 \,{\rm e}^{q_{\pi}    \left(1-\frac{r_{ij}}{a} \right)}  \,F_c(r_{ij}), \label{eq:tb0}\\
V_{pp\sigma}(r_{ij}) &=& \gamma_1 \,{\rm e}^{q_{\sigma} \left(1-\frac{r_{ij}}{a_1} \right)} \,F_c(r_{ij}) .
\label{eq:tb1}
\end{eqnarray}
where $a$ is the nearest neighbor distance within a layer, 
$a=1.418$\,{\rm \AA}, and $a_1$ is the
interlayer distance, $a_1=3.349$\,{\rm \AA}. 
First neighbors interaction in
a plane is taken equal to the commonly used value, $\gamma_0 = 2.7$\,eV.\cite{Castro09_RevModPhys}
Second neighbors interaction $\gamma_0'$ 
in a plane is set \cite{Castro09_RevModPhys}  to
$0.1\times\gamma_0$ 
which fixes the value of the ratio $q_{\pi}/{a}$ in equation (\ref{eq:tb0}).
The inter-layer coupling between two $p_z$ orbital in $\pi$ configuration 
is $\gamma_1$. $\gamma_1$ is fixed to obtain a good fit with ab-initio 
calculation around Dirac energy in AA stacking 
and AB bernal stacking  and then to get $\gamma_t=0.34$\,eV 
(figures \ref{Fig_AA_AB}, \ref{Fig_graph_mono_AA_AB} and \ref{Fig_DOS_graphene_LMTO_TB}) which results in 
$\gamma_1=0.48$\,eV.
We chose the same coefficient of the exponential decay 
for $V_{pp\pi}$ and $V_{pp\sigma}$,
\ben
\frac{q_{\sigma}}{a_1} ~=~ \frac{q_\pi}{a} 
~=~ \frac{{\ln} \left(\gamma_0'/\gamma_0 \right)}{a - a_0}
~=~ 2.218\,\Ang^{-1},
\een
with $a_0 = 2.456$\,{\rm \AA} the distance between second
neighbors in a plane. 
Intra-layer and inter-layer coupling terms $t(r_{ij})$ calculated with these
parameters are shown on figure \ref{Fig_couplageIntraInter}.
In (\ref{eq:tb0}) and (\ref{eq:tb1}), 
a smooth cut-off function is introduced,\cite{Mehl96}
\ben
\ds F_c(r) ~=~  \left( 1 + {\rm e}^{ \frac{r - r_c}{l_c} }  \right)^{-1},
\label{Eq_cutoffFunction}
\een
with $r_c$ the cut-off distance and 
$l_c= 0.5$\,a.u.\,$ = 0.265$\,{\rm \AA}.\cite{Mehl96}
For $r \ll r_c$, $F_c(r) \simeq 1$; and 
for $r \gg r_c$, $F_c(r) \simeq 0$.
All results presented in this article are calculated with
$r_c = 2.5 a_0 = 6.14$\,{\rm \AA}.
We have checked that the results are independent of $r_c$ for $r_c >4.9$\,{\rm \AA}.

All $p_z$ orbitals have the same on-site energy $\epsilon_i$ (equation
(\ref{Eq_hamilt})).
$\epsilon_i$ is set to $-0.78$\,eV so that
the energy $E_{\rm D}$ of the Dirac point is equal to zero in mono-layer
graphene.
$\epsilon_i$ is not zero because the intra-layer coupling between atoms
beyond first neighbors 
breaks the electron/hole symmetry and then shifts $E_{\rm D}$.


\begin{thebibliography}{99}

\bibitem{Wallace47}
P. R. Wallace, Phys. Rev. {\bf 71}, 622 (1947).
\bibitem{Castro09_RevModPhys}
A. H. Castro Neto, F. Guinea, N. M. R. Peres, K. S. Novoselov, and A. K. Geim,
{Rev. Mod. Phys.} {\bf 81}, 109 (2009).
\bibitem{Latil06}
S. Latil and L. Henrard, 
{Phys. Rev. Lett.} {\bf 97}, 036803 (2006).
\bibitem{MacdoABC}
F .Zhang, B. Sahu, H. Min, and A. H. MacDonald, 
Phys. Rev. B {\bf 82},  035409 (2010).
\bibitem{Varchon_prb08}
F. Varchon, P. Mallet, J.-Y. Veuillen, and L. Magaud, 
{ Phys. Rev. B} {\bf 77}, 235412 (2008).
\bibitem{Ohta06}
T. Ohta, A. Bostwick, T. Seyller, K. Horn, and E. Rotenberg,
Science {\bf 313}, 951 (2006).
\bibitem{Coletti10}
C. Coletti, C. Riedl, D. S. Lee, B. Krauss, L. Patthey, K. von Klitzing, J. H. Smet, and U.Starke, 
Phys. Rev. {\bf 81}, 235401 (2010).
\bibitem{Brihuega08} 
I. Brihuega, P. Mallet, C. Bena, S. Bose1, C. Michaelis, L. Vitali, F. Varchon, L. Magaud, K. Kern, and J. Y. Veuillen, 
Phys. Rev. Lett. {\bf 101}, 206802 (2008).
\bibitem{Hass06}
J. Hass, R. Feng, T. Li, X. Li, Z. Zong, W. A. de Heer, P. N. First, E. H. Conrad, C. A. Jeffrey, and C. Berger,
Appl. Phys. Lett {\bf 89}, 143106 (2006).
\bibitem{Hass08}
J. Hass, W. A. de Heer, and E. H. Conrad,
J. Phys: Condens. Matter {\bf 20}, 323202 (2008).
\bibitem{Emtsev08}
K. V. Emtsev, F. Speck, Th. Seyller, L. Ley, and J. D. Riley, Phys. Rev. B {\bf 77}, 155303 (2008).
\bibitem{Sprinkle09}
M. Sprinkle, D. Siegel, Y. Hu, J. Hicks, A. Tejeda, A. Taleb-Ibrahimi, P. Le F\`evre, F. Bertran, 
S. Vizzini, H. Enriquez, S. Chiang, P. Soukiassian, 
C. Berger, W. A. de Heer, A. Lanzara, and E. H. Conrad,
Phys. Rev. Lett. {\bf 103}, 226803 (2009).
\bibitem{Hicks11}
J. Hicks et al Phys. Rev. B {\bf 83}, 205403 (2011).

\bibitem{Hass_prl08}
J. Hass, F. Varchon, J. E. Millan-Otoya, M. Sprinkle, N. Sharma, W. A. de Heer, C.Berger, P. N. First, L. Magaud,
and E. H. Conrad, 
{ Phys. Rev. Lett.} {\bf 100}, 125504 (2008).
\bibitem{Andrei_Ni11}
A. Luican, G. Li, A. Reina, J. Kong, R. R. Nair, K. S. Novoselov, A. K. Geim, and E. Y. Andrei, 
{Phys. Rev. Lett.} {\bf 106}, 126802 (2011).
\bibitem{Andrei_graphite10}
G. Li, A. Luican, L. M. B. Lopes dos Santos, A. H. Castro Neto, A. Reina, J. Kong, and E. Y. Andrei, 
{Nat. Phys.} {\bf 6}, 109, (2010).


\bibitem{Miller09}
D. L. Miller, K. D. Kubista, G. M. Rutter, M. Ruan, W. A. de Heer, P. N. First, and J. A. Stroscio,
Science {\bf 324}, 924 (2009).

\bibitem{Berger06}
C. Berger, Z. M. Song, X. B. Li, X. S. Wu, N. Brown, C. Naud, D. Mayou, T. B. Li, J. Hass, 
A. N. Marchenkov, E. H. Conrad,
P. N. First, and W. A. de Heer,
Science {\bf 312}, 1191 (2006).
\bibitem{Sadowski06}
M. L. Sadowski, G. Martinez, M. Potemski, C. Berger, and W. A. de Heer, Phys. Rev. Lett. {\bf 97}, 266405 (2006).

\bibitem{Varchon_stm08}
F. Varchon, P. Mallet, L. Magaud, and J.-Y. Veuillen, Phys. Rev. B {\bf 77}, 165415 (2008).
\bibitem{Rong93} 
Z. Y. Rong, and P. Kuiper, Phys. Rev. B {\bf 48}, 17427 (1993).




\bibitem{Dossantos07}
J. M. B. Lopez dos Santos, N. M. R. Peres, and A. H. Castro Neto,
Phys. Rev. Lett. {\bf 99}, 256802 (2007).

 \bibitem{CastroNeto_condmat} 
J. M. B. Lopes dos Santos, N. M. R. Peres, and A. H. Castro Neto, CondMat: ArXiv.1202.1088v1 (2012).

\bibitem{Latil07} 
S. Latil, V. Meunier, and L. Henrard, {Phys. Rev. B} {\bf 76}, 201402(R), (2007).
\bibitem{Shallcross08}
S. Shallcross, S. Sharma, and O. A. Pankratov, 
Phys Rev. Lett. {\bf 101}, 056803 (2008).
\bibitem{Mele}
E. J. Mele, Phys. Rev. B {\bf 81}, 161405 (R) (2010).

\bibitem{Mele2}
E. J. Mele, Phys. Rev. B {\bf 84}, 235439 (2011).
\bibitem{Suarez}
E. Suarez Morell, J. D. Correa, P. Vargas, M. Pacheco, and Z. Barticevic, Phys. Rev. B {\bf 82}, 121407(R) (2010).
\bibitem{Bistritzer_PNAS}
R. Bistritzer and A. H. MacDonald, Proc. Natl. Acad. Sci. USA {\bf 108}, 12233 (2011).
\bibitem{Bistritzer2}
R. Bistritzer and A. H. MacDonald, Phys. Rev. B {\bf 81}, 245412 (2010). 
\bibitem{Trambly10}
 G. Trambly de Laissardi\`ere, D. Mayou, and L. Magaud, {\it Nano Lett.} {\bf 10}, 804 (2010).
 
\bibitem{Veuillen}
J.-Y. Veuillen et al., private communication.
\bibitem{Campanera}
J. M. Campanera, G. Savini, I. Suarez-Martinez, and M. I. Heggie,
Phys. Rev. B {\bf 75}, 235449 (2007).

\bibitem{LDA}
These calculations where performed with VASP with LDA which gives a better description of the interlayer distance.

\bibitem{note}
Let us note that other bands contributes also to the central peak of the density of states  but do not give linear 
dispersion close to the Dirac points.


\bibitem{angles}
In Fig.\,\ref{Fig_vitesse}, the velocity (orange line) is plotted  as a
function of the angle $\theta$ and not $\alpha$
as in Ref. \onlinecite{Bistritzer_PNAS}.
For small angles:  $\alpha={C}/{\theta}$, where  $C = \gamma_t / (\hbar
V_{\rm mono} K_{\rm D})$.
In Ref. \onlinecite{Bistritzer_PNAS},
parameters are deducted from experiment and lead to $C=0.621$.
We use a different TB scheme, fitted on ab-initio calculations that
underestimate velocity value.
Therefore to compare our results with those in Ref.
\onlinecite{Bistritzer_PNAS}, we use $C = 0.67$ to
obtain the same angle value for the first magic angle.
This last $C$ value is used to plot the orange variation.



\bibitem{Berger93}
C. Berger, E. Belin, and D. Mayou,
{Annale de Chimie-Sciences des Materiaux} {\bf 18}, 485 (1993).


\bibitem{Triozon02} F. Triozon, Julien Vidal, R. Mosseri, and Didier Mayou,
{Phys. Rev. B} {\bf 65}, 220202 (2002). 


\bibitem{Trambly06} G. Trambly de Laissardi\`ere, J.-P. Julien, and D. Mayou,
{Phys. Rev. Lett.} {\bf 97}, 026601 (2006). 


\bibitem{SlaterKoster54}
J. C. Slater and G. F. Koster, 
{ Phys. Rev.}  {\bf 94}, 1498
(1954). 

\bibitem{relax}
The atomic positions of the (6,7) bilayer have been fully relaxed using 
VASP and the LDA functionnal without any noticeable effect either on the geometry 
-no buckling- nor on the electronic structure.

\bibitem{LMTO}
O.K. Andersen Phys. Rev. B {\bf 12}, 3060 (1975).

\bibitem{Mehl96}
M. J. Mehl and D. A. Papaconstantopoulos, Phys. Rev. B {\bf 54}, 4519 (1996).

\end{thebibliography}
\end{document}